\documentclass[preprint]{scrartcl}

\usepackage{comment}
\usepackage{hanging}
\usepackage{epsfig}
\usepackage[utf8]{inputenc}
\usepackage{textgreek}
\usepackage{threeparttable}
\usepackage{booktabs}
\usepackage{amssymb}
\usepackage{amsfonts}
\usepackage{amsthm}
\usepackage{graphics} 
\usepackage{amsmath} 
\usepackage{subfigure}
\usepackage{graphicx}
\usepackage{lscape}
\usepackage{multicol} 
\usepackage{graphicx, epstopdf}
\usepackage[table,xcdraw]{xcolor}
\usepackage{caption}
\usepackage{xspace,epsfig,url}
\usepackage{enumerate}
\usepackage{float}
\usepackage{epsf}
\usepackage{psfrag}
\usepackage{color,soul}
\usepackage{bm}
\usepackage{comment}
\usepackage{cases}
\usepackage{hyperref}
\usepackage{amssymb}
\usepackage{array} 

\usepackage[toc,page]{appendix}
\usepackage{tabularx}
\usepackage{threeparttablex}

\graphicspath{ {./figures/} }

\begin{document}





\title{Gamification Box: Automated Personalized Rewards for Improved Task Experience}



\author{
    Ali Utku Ozaslan\textsuperscript{*}%
    \thanks{Computer Engineering Department,
    Kadir Has University, Istanbul, Turkey.\\
    \textsuperscript{*}Corresponding author:
    \texttt{aliutku.ozaslan@stu.khas.edu.tr}}
    \and
    Sadra Sadraddini%
    \thanks{Sony Research, Boston, MA, USA}
    \and
    Deniz Marti%
    \thanks{California State University,
    Northridge, CA, USA}
}

\date{}

\maketitle

\begin{abstract}

We introduce a gamified rewarding system, called ”Gamification Box”, and conduct a user study to examine its effects on task experience and performance. Gamification Box provides an interface for automated rewards that are intrinsically enjoyable activities, such as watching short video clips. In a controlled user experiment, 118 participants completed a short-term memory task involving meaningless syllables under either a gamified reward condition or a non-rewarded control condition. Participants in the gamified condition reported significantly greater perceived task enjoyment than those in the control group. Importantly, this increase in perceived enjoyment did not come at the cost of performance: no significant differences were observed in free recall or recognition accuracy between conditions. These findings suggest that while gamified rewards may not directly enhance immediate performance, they function as a motivational regulator, thereby improving task enjoyment. The primary benefit of such systems may therefore lie in supporting sustained effort over time, where cumulative performance gains are more likely to emerge.


\end{abstract}







\section{Introduction}
\label{sec:intro}

Many educational and work environments require individuals to engage in cognitively demanding tasks that are perceived as boring, repetitive, or intrinsically uninteresting. Such a gap between task demands and intrinsic motivation can result in procrastination and disengagement, often leading to frustration, reduced well-being, and suboptimal learning outcomes, as deadlines and responsibilities accumulate (Beutel et al., 2016; Steel \& Klingsieck, 2016). While many individuals recognize the long-term value of sustained effort, maintaining attention and endurance during such tasks remains a persistent challenge.

Traditional education systems encourage desirable behaviors by awarding grades and ranking students on leaderboards. This system is effective for certain types of learners who are motivated by high marks and the competitive aspect of outperforming their peers (Smiderle et al., 2020). This motivation can be understood through the lens of social comparison theory, which suggests that individuals evaluate their own abilities in relation to others (Garcia et al., 2013). However, not all students respond to such abstract incentives. Some learners demonstrate greater engagement when tasks are inherently enjoyable, rather than when motivations rely on external validation through scores or rankings. For such learners, maintaining sustained attention often depends less on abstract rewards and more on how effort and relief are structured over time.

One widely adopted approach to supporting sustained focus is the Pomodoro Technique, a time management strategy that alternates fixed study intervals with short breaks that are typically 25 minutes of work followed by 5 minute breaks (Biwer et al., 2023). This temporal structure at work and rest reduces cognitive fatigue and, hence, improves task persistence. However,this approach does not automate the transition between study and leisure activities, requiring users to manually initiate each switch and rely on self-control. Gamification, on the other hand, aims to bring game paradigms into non-game contexts (Marczewski, 2013). Gamification systems commonly include Idle Games, Role-playing Games (RPGs), non-RPGs, or hybrid approaches that combine elements of both genres (Caponetto et al., 2014).

In this paper, we introduce the Gamification Box (GB), which is a non-RPG gamification approach that uses an automation mechanism between studying and getting rewards. GB’s design draws on the principles of operant conditioning (Skinner, 2019), in which behavior is shaped through contingent and immediate positive reinforcement. Traditional operant conditioning research has often focused on tangible reinforcers such as food or toys (Laule, 2010). In contrast to RPG approaches, GB does not rely on symbolic in-game rewards or avatars. It employs intrinsically meaningful activities such as enjoyable videos as rewards. Therefore, our approach emphasizes reinforcement that aligns with users’ momentary desires rather than symbolic or delayed incentives.

The remainder of the paper is structured as follows. In Sec. 2, we review the related work. In Sec. 3, the specifications of GB software are explained. Next, we explain the experimental design method in Sec. 4 and report the results of the study in Sec. 5. Finally, in Sec . 6, we discuss our findings and their relation to high level paradigms and conclude with potential future directions in Sec. 7.

\section{Background and Related Work} 
\label{sec:background}

\subsection{Pomodoro Technique}
The Pomodoro technique is one of the most well-known methods of studying and learning (Cirillo, 2018). It imposes a regular rhythm of effort and rest, where the individual studies for 25 minutes, then takes a 5-minute break. This continues until the fourth study session. After the fourth session, the individual takes a 15-30-minute break. As reported by many studies, the Pomodoro technique has benefits for learning and studying (Biwer et al., 2023). For example, (Septiani et al., 2020) reported that an experimental group that used the Pomodoro technique had a significantly higher improvement in writing skills compared to the control group that did not use it.

The literature on retention intervals consistently demonstrates that spaced learning is more effective than massed learning (Godbole et al., 2014; Goossens et al., 2012; Vlach et al., 2008). Numerous mobile and computer-based applications have been developed to support spaced learning; however, many of these systems rely on similar algorithmic structures and require users to manually manage transitions between study and break activities. As a result, these tools demand continuous self-regulation in moments of cognitive fatigue. To address this limitation, our proposed system, GB, automates the transition between effort and reward, which helps reduce the reliance on user-initiated task switching.

\subsection{Gamification Studies}
Gamification can be described as a behavioral delivery system for spaced learning (distributing learning over time), keeping learners returning at optimal intervals rather than cramming (learning packed in one session). A substantial body of research on retention intervals demonstrates that spaced learning leads to superior memory performance compared to massed learning across a variety of populations and task designs (Godbole et al., 2014; Goossens et al., 2012; Murphy et al., 2023; Yuan, 2022). For example, Vlach et al.( 2008) conducted a study on three-year-old children on their learning performance. The duration of study intervals was 10 seconds, while the duration of break intervals was kept at 30 seconds. They showed those who were in the spaced learning condition had a significantly higher performance than those who were in the massed learning condition.

In contrast, evidence regarding the impact of gamification on performance outcomes is more mixed. For example, Khoshnoodifar et al. (2023) found no significant performance differences between intervention and control groups when using common gamification elements such as scoreboards, points, levels, objectives, and progress indicators. These mechanics are widely adopted in mainstream gamification systems (Hakulinen et al., 2013). Despite limited and inconsistent effects on performance, numerous studies report that such gamification techniques reliably increase motivation and engagement compared to non-gamified conditions (Denny, 2013; Farzan et al., 2008).
A growing body of research suggests games themselves can improve certain cognitive and behavioral skills that could be useful in other areas of life such as education and learning (Boyle et al., 2014). The authors Qiao et al. (2024) explored the effects of gamified competition and cooperation on success. They showed that the competitive condition group had a significantly higher success rate in morphological learning than the cooperative condition group.

As attested by Chen and Liang (2022), gamification of lecture material significantly increases student engagement with study material. In a study that was conducted on a programming course, the students were split into two groups. A group was rewarded with points, badges and a rank on the class leaderboard. The other group was non-rewarded, and the findings showed that there was a significant difference between the reward condition group and the no reward condition group. The reward condition group had a significantly higher performance than the no reward condition group (Smiderle et al., 2020). Guin et al. (2012) has also demonstrated similar findings.

\subsection{Related Gamification Software Platforms}
There are numerous gamification platforms available. Among those platforms, perhaps one of the prominent ones is Kahoot, which is a test-driven platform where individuals gain points from answering questions. This serves as a game mechanic for competition among students. Kahoot and its effects on learning and engagement have been investigated by researchers (Licorish et al., 2018; Wang \& Tahir, 2020; Zhang \& Yu, 2021).

Many productivity apps use aspects of gamification. Examples include Habitica which highly relies on RPG elements to reinforce studying by the completion of pre-determined tasks that are manually defined by the users, and Forest, which is based on the Pomodoro technique and specific characteristics of the common Idle Games. In Pomofarm, which is also based on the Pomodoro technique (Cirillo, 2018), the players study for a pre-set duration, they earn in-game currencies (vegetable, in this case) that grow with each study session. During the break period, the player gets to sell their vegetables, earn more coins, and plant more vegetables. A major problem with the existing software is that it may not attract individuals who do not find the accumulation of numbers entertaining. For a stimulus to be rewarding, one needs to feel an intrinsic motivation.

\section{Proposed System: The Gamification Box (GB)}
\label{sec:gamificationbox}
\label{sec:materials}
GB is ultimately intended to be used to gamify the experience of engaging in cognitive tasks.

It is available for download at the \href{https://ali-utku-ozaslan.itch.io/gamification-box}{Gamification Box project page}.

\subsection{Gamification Box Mechanics}
GB has options for setting study/work and break modes. The main menu is shown in Figure 1. The user can choose the study mode from the options shown on the top left. Next, the user enters a study/work mode. An example is shown in 2, where an instance of GB is on the top left of the screen showing the total study time (since the start of the day) and a countdown of the current study or the reward period. It also gives the chance to switch the reward period early and carry that amount to the next reward session. However, a reward period cannot be started before the completion of a study one. In addition, it is possible to pause the whole software and resume later. The full description of the available options and related features are available on the website, and here we only describe the high level ones.
\begin{figure}[t]
    \centering
    \includegraphics[width=0.9\linewidth]{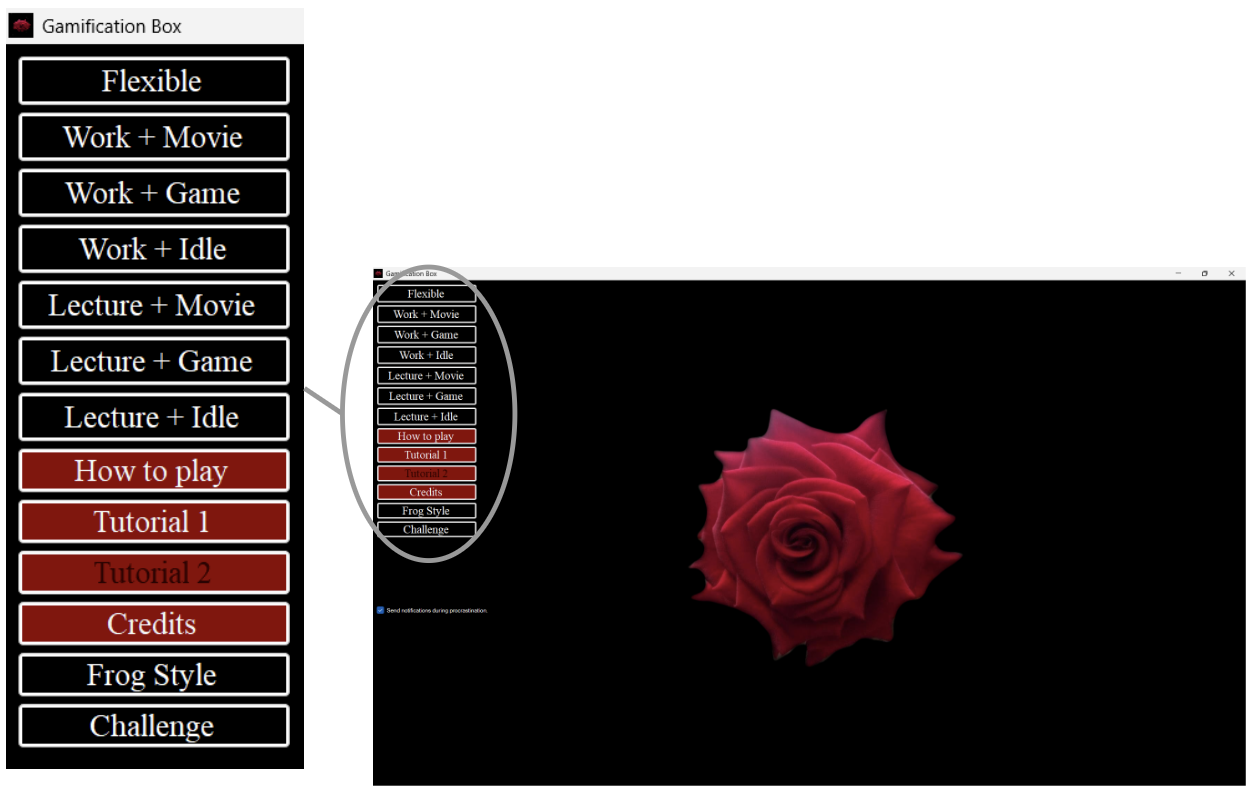}
    \caption{Gamification Box Main Menu}
    \label{fig_gb_main}
\end{figure}
\begin{figure}[t]
    \includegraphics[width=0.95\linewidth]{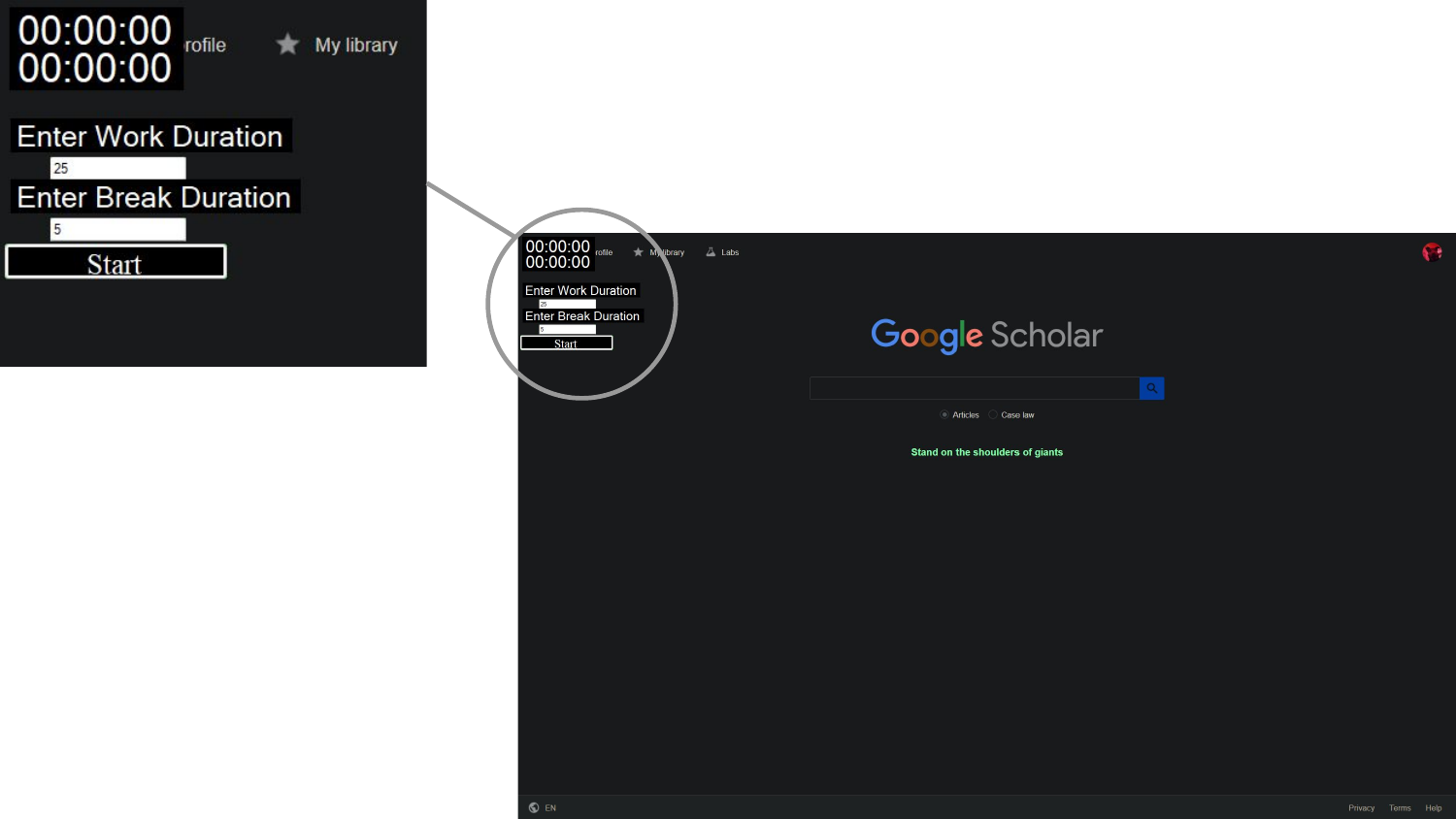}
    \caption{Example of a page view to start a session with Gamification Box}
    \label{fig_gb_work}
\end{figure}

The Flexible mode gives the user the ability to study or work without interruption while accumulating reward time to use for breaks when they desire to take a break. The reward is given according to the ratio that the user has set before starting the study session. For instance, if a 2/1 ratio is chosen, a 30 minute studying period will be followed by a 15 minute reward period. Alternatively, the user can give their desired study duration and break duration, as illustrated in Figure 2. The reward can be of any nature that the user wants: watching a movie, playing a video game, or an idle activity, such as reading a book. Personalization is key, as the perceived value of a reward depends on individual preferences.

The other modes are more precise in their descriptions as they are combinations of different pre-defined study categories (i.e., Lecture or Work) and reward categories (i.e., Movie, Game or Idle). There is a mode available for each possible combination of Study+Reward categories.

Different categories are to account for different needs for different study and reward types while using the application. For example, the category Lecture is for watching a lecture video and as such it is differentiated from Work by the inclusion of an automated stop, in the sense that GB pauses the lecture media player. Similarly, GB automates pausing the Movie period. For the modes that have Game as the reward, a countdown is voiced to alert the user of the upcoming switch to study period so that they can pause their game in a timely manner and not lose progress. The modes with the Idle reward, on the other hand, do not include any of these additional features since the user is expected to be away from the keyboard when using this reward type.

In tasks that are not intrinsically motivating, a reward may offer an external incentive to complete these tasks. Automating transitions between task and reward is designed to reduce user burden of task switching. The user does not need to press more keys than a simple hotkey if necessary to see the reward take place or continue the study session from where they left, with their reward not being reset to zero. Preserving unused reward time makes the experience non-punitive.

\section{Experiment}
\label{sec:methods}
A user experiment was conducted to assess the impact of gamification on user experience and task performance in activities that are not intrinsically interesting or rewarding. The experiment was carried out using a limited version of the GB software, specifically tailored for the purposes of this study. We describe the details of the experiment starting with participant selection in Section 4.1, followed by the experimental procedure in Section 4.2. Finally the methods used in analyzing the experiment outcomes are presented in Section 4.3.

\subsection{Participants}
\label{sec:participants}

Participants were recruited from a university and included undergraduate students, graduate students, and university staff. The study protocol was approved by the Institutional Review Board and written informed consent was obtained from all participants, who were 18 years old or older. Previous research has shown that age can significantly influence various cognitive functions, including memory (Mazurek et al., 2015; Rhodes et al., 2019). Therefore, efforts were made to minimize age variability within the sample. Participants did not receive monetary compensation. However, some courses offered additional course credit or points as an incentive for participation. All attendees signed an informed consent.

An a priori power analysis conducted using G*Power indicated that a sample size of 126 participants would be required to detect a medium effect size with 80\% statistical power. Enrollment continued until this target sample size was reached.

\subsection{Methods}
\label{sec:procedure}
The goal was to evaluate the impact of the personalized rewards on engagement with tasks that are otherwise considered boring or uninteresting. The experimental task involved memorizing a set of meaningless syllables presented to participants during the study. This is a commonly used experimental technique that is used to study short-term memory and task engagement under tightly controlled conditions, because meaningless syllables reduce the influence of semantic meaning, prior familiarity, and individual differences in background knowledge (Ebbinghaus, 1949; Cowan, 2008; Aben et al., 2012).

After receiving instructions, participants were randomly assigned to the Control or Reward group The experiment consisted of three 1-minute study sessions interspersed with two 1-minute breaks, for a total duration of 5 minutes. A schema of this process is provided in Figure 3. During each study session, participants were presented with meaningless syllables (e.g., Zob) sequentially, one at a time. After each syllable, a fixation cross appeared on the screen to provide clear separation before the next syllable was presented (see Figure 4). A total of 20 syllables were presented in each study session, with each syllable appearing three times over the course of the experiment. The full list of syllables is provided in the Appendix. During the two 1-minute breaks between study sessions, participants in the Reward group watched a YouTube video of their selection. In contrast, participants in the Control group viewed a public service advertisement. Example moments of a participant going through the study and break phases of the experiment are shown in Figure 5.

Following the experiment, each participant completed a distractor task in which they counted aloud in increments of three, starting from zero, for 30 seconds. Afterward, they were instructed to type all the syllables they could remember using the keyboard (free recall test). Next, they completed a recognition test consisting of 40 Yes/No questions, in which they indicated whether each syllable had been included in the list presented during the experiment. Finally, participants completed a self-report questionnaire consisting of 12 items rated on a 5-point Likert scale (1 = strongly disagree, 5 = strongly agree). The questionnaire assessed multiple dimensions of task experience, including perceived enjoyment, perceived effort, boredom, focus, temptation, and perceived performance (see Appendix for the full questionnaire).
\begin{figure}[H]
\centering
  \includegraphics[width=\linewidth]{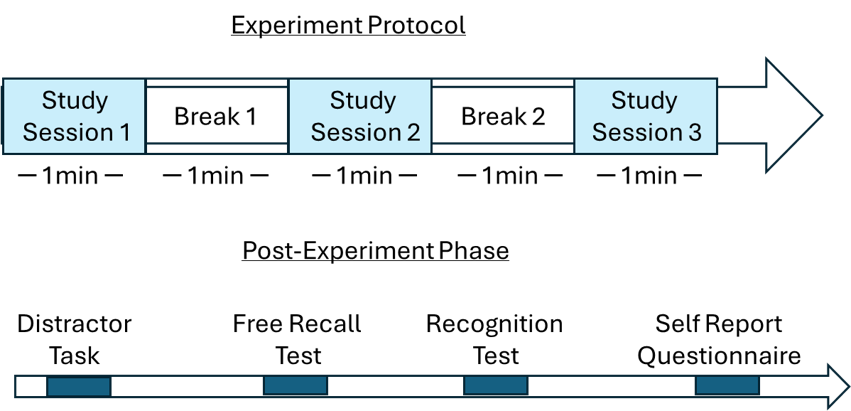}
  \caption{Overview of the experimental design. Participants completed three 1-minute study sessions interleaved with 1-minute breaks, followed by a post-experiment phase including a distractor task, free recall, recognition testing, and a self-report questionnaire.}
  \label{fig:flowchart}
  \centering
\end{figure}
As detailed earlier, several methodological features were implemented to ensure experimental control. The distracting task reduced recency effects prior to recall. The use of meaningless syllables minimized semantic associations and ensured stimulus novelty across participants. Each syllable was presented for two seconds and preceded by a one-second fixation cross to support attentional focus. Consequently, these design choices allowed us for a controlled experimentation of motivational and performance-related outcomes under fixed spacing conditions.

\subsection{Data Analysis}
\label{sec:measures}
Data were analyzed using Jamovi (version 2.7.12). Recall performance was defined as the number of correctly recalled syllables during the free recall test. Recognition performance was defined as the proportion of correct responses in the Yes/No recognition test. Task tolerability, engagement and perceived enjoyment were assessed using self-report questionnaire responses measured on a Likert scale.

The independent variable was the presence of gamification (i.e. Reward vs Control group), and the dependent variables included recall accuracy, recognition accuracy, and self-reported Likert scale scores to each question separately.
\begin{figure}[H]
	\centering
        \subfigure[\label{fig:fixation}]%
	{\includegraphics[height=4cm]{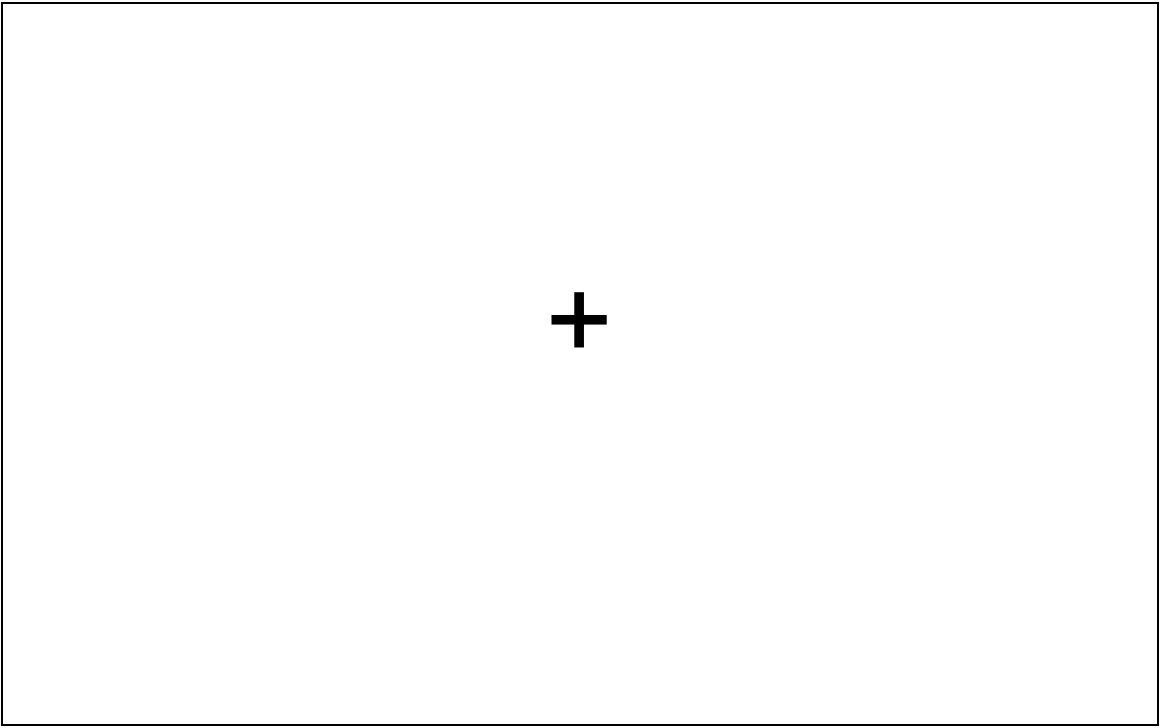}}
        \subfigure[\label{fig:meaninglessSyllable}]%
        {\includegraphics[height=4cm]{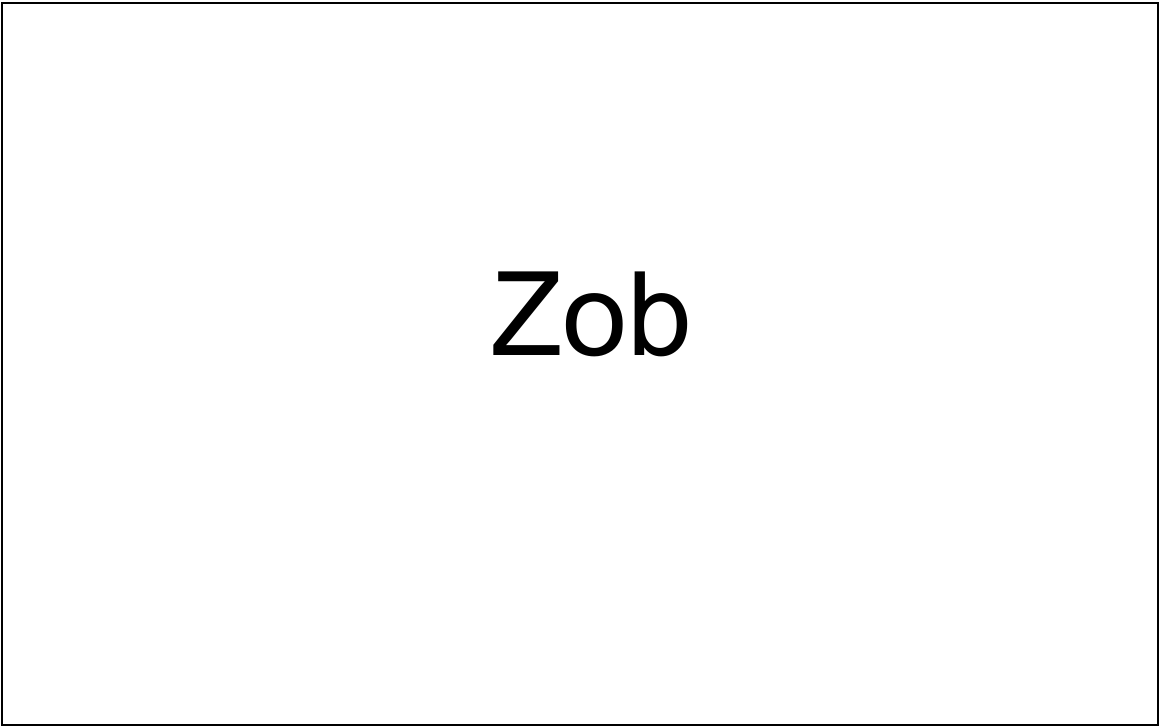}}
	\caption{[Left] Fixation Cross as attention grabber graphical content (Hirai et al., 2016). [Right] One of the meaningless syllables used in the experiment.}
	\label{fig:application_layout}
\end{figure}
Perceived enjoyment, defined as the primary subjective outcome, was connected to participants’ responses to the questionnaire item, “It was fun doing the task of the experiment.” The responses were treated as continuous variables for statistical analysis. Descriptive statistics were calculated for all dependent variables. Independent samples t-tests were used to compare self-reported questionnaire outcomes and performance measures between the reward and control conditions. Statistical significance was assessed at an α level of .05. Non-significant questionnaire items are reported descriptively, with full item statistics provided in the Appendix.
\begin{figure}[H]
\centering
  \includegraphics[width=\linewidth]{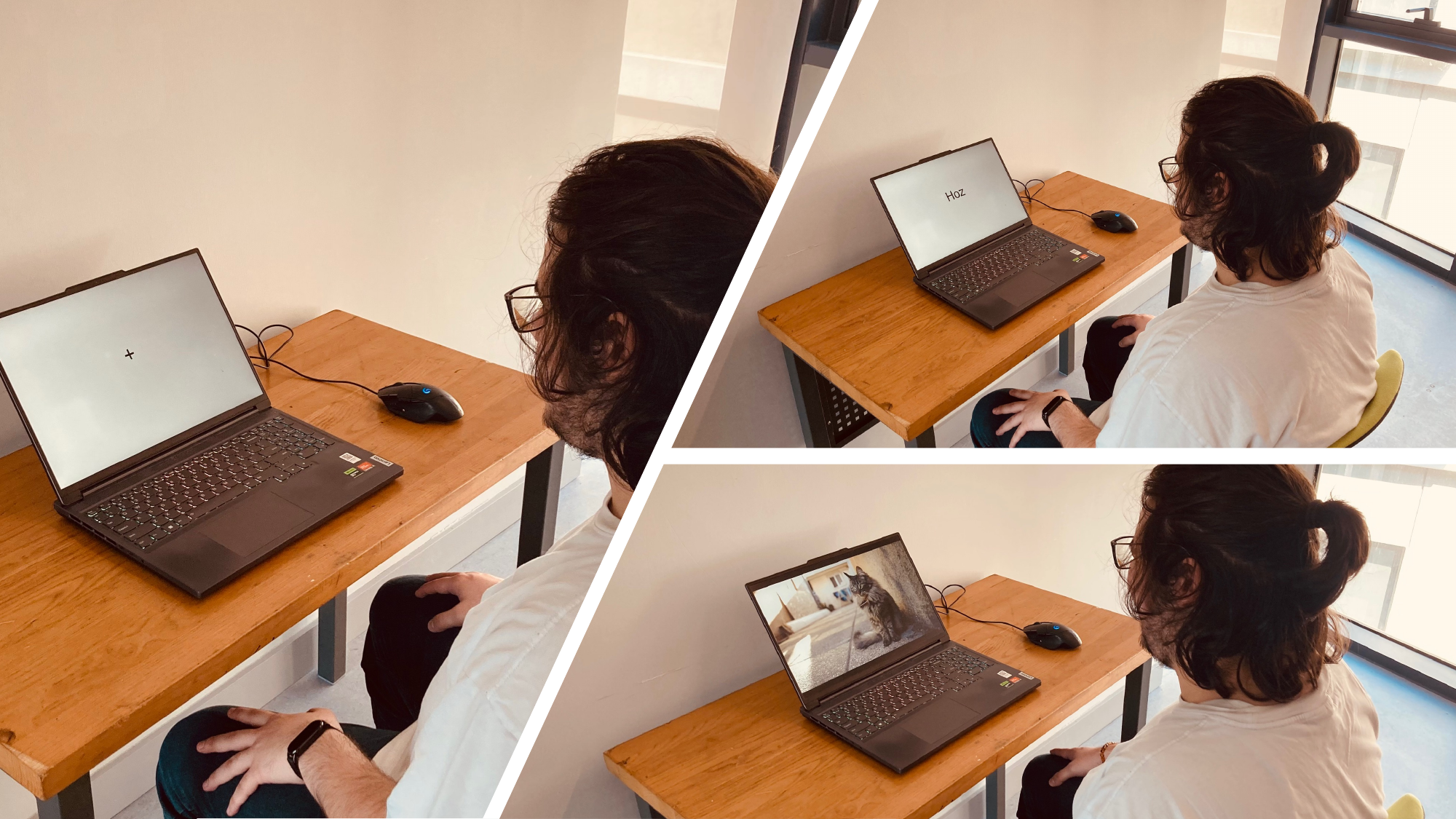}
  \caption{Example participant in the Reward condition. The participant views a fixation cross [Left] to orient attention, followed by presentation of a meaningless syllable to encode [Top right]. During the subsequent break phase, the participant is rewarded with a self-selected video [Bottom right].}
  \label{fig:participant}
  \centering
\end{figure}

\section{Results}
\label{sec:results}

A total of 128 participants participated in the study. Data from the first 10 participants were excluded because they were collected using an earlier version of the experimental protocol that did not include the public advertisement during break periods. The final sample consisted of 118 participants.

Perceived Enjoyment: Participants in the Reward condition reported significantly higher perceived enjoyment in response to the item “It was fun to do the task at hand” compared to participants in the Control condition, t(116) = 2.39, p = .019, d = 0.4396. The mean perceived enjoyment rating was 3.90 (SD = 1.08) in the Reward condition and 3.45 (SD = 0.99) in the Control condition (see Figure 6).

Performance outcomes: Free recall performance did not differ significantly between conditions. Participants in the Reward condition recalled an average of 8.24 syllables (SD = 3.33), compared to 7.61 syllables in the Control condition (SD = 3.32), t(116) = 1.035, p = .303, d = 0.1909. Similarly, recognition performance was comparable between the Reward condition (M = 33.03, SD = 4.44) and the Control condition (M = 33.29, SD = 4.24), t(116) = -0.316, p = .752, d = -0.0583, indicating no significant difference in recognition accuracy between groups. The percentage of correctly recalled and recognized syllables is visualized in Figure 6, illustrating the performance distributions across conditions. 
\begin{figure}[H]
\centering
  \includegraphics[width=\linewidth]{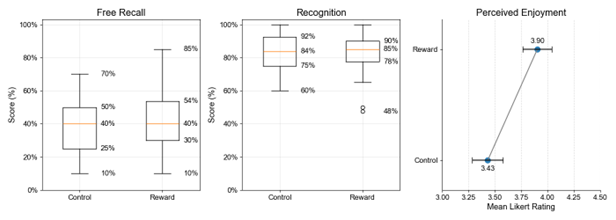}
  \caption{Comparison of outcomes between the Reward and Control conditions. [Left] Free Recall Score; [Middle] Recognition Score; [Right] Enjoyment Score, shown as mean Likert ratings with error bars.}
  \label{fig:figure6}
  \centering
\end{figure}
Other self-report measures: No statistically significant differences were observed between the Reward and Control conditions in perceived effort, boredom, focus, temptation, or perceived performance. Detailed results are provided in Appendix Table A5..




\section{Discussion}
\label{sec:discussion}
This study explored the extent to which gamified reinforcement affects task performance and perceived enjoyment. In a short memorization task, we found no statistically significant differences in memory performance between the reward and no-reward conditions. However, participants in the reward condition reported significantly higher perceived enjoyment of the task. This result suggests that GB can enhance the subjective experience of cognitively demanding tasks without impairing immediate performance.

While one might expect gamification to improve both performance and perceived enjoyment, the absence of performance differences is theoretically consistent with prior research showing that memory outcomes are primarily driven by factors such as spacing and repetition, which were held constant across conditions in the present design. Because these core learning parameters were controlled, improvements in perceived enjoyment were not expected to translate into enhanced memorization performance within this short task.

One possible implication of increased task enjoyment is that, over longer periods of use, users may be more willing to remain engaged with the task, thereby increasing exposure and potentially improving performance. However, this possibility was not tested in the present study and future research should examine the effects of longer periods of use. It is also possible that repeated pairing of effortful tasks with enjoyable breaks could change how users experience the task over time, which could make the task feel less aversive. This possibility warrants further investigation, particularly to determine whether repeated use of such systems leads to changes in users’ task perception.

A key limitation of our study was its short duration, which did not allow observing potential outcomes for tasks with a longer duration. Significantly increased perceived enjoyment that we observed in this study may reflect greater tolerance for task-related boredom and cognitive fatigue, suggesting that participants in the Reward condition may be more likely to persist over extended study periods. Additionally, the task of the experiment, memorizing meaningless syllables with no reward or penalty for individual performance, does not fully reflect real-world learning contexts, where successful completion often depends on achieving a certain level of performance. Future work should therefore examine whether performance differences emerge in settings that involve sustained engagement and meaningful incentives, such as performance-contingent monetary rewards.

Finally, as a study, our experiment focused solely on automated rewards within a highly controlled and minimally personalized setting. Future research could explore richer forms of personalization, more flexible study schedules, and alternative reward types (e.g., interactive or game-based rewards) to better reflect real-world usage.

More broadly, our study is a step towards understanding how we can improve the learning experience for individuals. Learning often requires individuals to tolerate discomfort, effort, and delayed gratification. When tasks lack intrinsic appeal, the reliance on distant or abstract rewards, external motivational factors such as grades or credentials, may be insufficient to sustain engagement. Automated rewarding, on the contrary, provides immediate relief and satisfaction because it reduces the need for self-control. This mechanism may be particularly relevant for individuals who struggle to persist through monotonous, tedious, or compulsory tasks.

\section{Conclusion}
\label{sec:conclusion}
Our findings demonstrate that personalized rewards enhance the subjective experience of tasks that otherwise are not interesting without negatively affecting cognitive performance for tasks of short duration. In this study, Gamification Box appeared to function as a support mechanism that improves task enjoyment rather than acting as an instant performance enhancer. Although short-term cognitive performance gains were not observed, enhanced perceived enjoyment could become consequential in contexts where sustained engagement is required. 

\section*{References}
Aben et al., 2012. Aben, B., Stapert, S., and Blokland, A. (2012). About the distinction between working memory and short-term memory. \textit{Frontiers in psychology}, 3:301.

Beutel et al., 2016. Beutel, M. E., Klein, E. M., Aufenanger, S., Brähler, E., Dreier, M., Müller, K. W., Quiring, O., Reinecke, L., Schmutzer, G., Stark, B., et al. (2016). Procrastination, distress and life satisfaction across the age range–a german representative community study. \textit{PloS one}, 11(2):e0148054.

Biwer et al., 2023. Biwer, F., Wiradhany, W., oude Egbrink, M. G., and De Bruin, A. B. (2023). Understanding effort regulation: Comparing 'pomodoro' breaks and self-regulated breaks. \textit{British Journal of Educational Psychology}, 93:353–367.

Boyle et al., 2014. Boyle, E. A., MacArthur, E. W., Connolly, T. M., Hainey, T., Manea, M., Kärki, A., and Van Rosmalen, P. (2014). A narrative literature review of games, animations and simulations to teach research methods and statistics. \textit{Computers \& Education}, 74:1–14.

Caponetto et al., 2014. Caponetto, I., Earp, J., Ott, M., et al. (2014). Gamification and education: A literature review. In \textit{European conference on games based learning}, volume 1, pages 50–57.

Chen and Liang, 2022. Chen, J. and Liang, M. (2022). Play hard, study hard? the influence of gamification on students' study engagement. \textit{Frontiers in Psychology}, 13:994700.

Cirillo, 2018. Cirillo, F. (2018). \textit{The Pomodoro Technique: The Acclaimed Time Management System That Has Transformed How We Work}. Crown Currency.

Cowan, 2008. Cowan, N. (2008). What are the differences between long-term, short-term, and working memory? \textit{Progress in brain research}, 169:323–338.

Denny, 2013. Denny, P. (2013). The effect of virtual achievements on student engagement. In \textit{SIGCHI conference on human factors in computing systems}, pages 763–772.

Ebbinghaus, 1949. Ebbinghaus, H. (1949). \textit{Experiments in memory}.

Farzan et al., 2008. Farzan, R., DiMicco, J. M., Millen, D. R., Brownholtz, B., Geyer, W., and Dugan, C. (2008). When the experiment is over: Deploying an incentive system to all the users. In \textit{symposium on persuasive technology}. Citeseer.

Garcia et al., 2013. Garcia, S. M., Tor, A., and Schiff, T. M. (2013). The psychology of competition: A social comparison perspective. \textit{Perspectives on psychological science}, 8(6):634–650.

Godbole et al., 2014. Godbole, N. R., Delaney, P. F., and Verkoeijen, P. P. (2014). The spacing effect in immediate and delayed free recall. \textit{Memory}, 22(5):462–469.

Goossens et al., 2012. Goossens, N. A., Camp, G., Verkoeijen, P. P., Tabbers, H. K., and Zwaan, R. A. (2012). Spreading the words: A spacing effect in vocabulary learning. \textit{Journal of Cognitive Psychology}, 24(8):965–971.

Guin et al., 2012. Guin, T. D.-L., Baker, R., Mechling, J., and Ruyle, E. (2012). Myths and realities of respondent engagement in online surveys. \textit{International Journal of Market Research}, 54(5):613–633.

Hakulinen et al., 2013. Hakulinen, L., Auvinen, T., and Korhonen, A. (2013). Empirical study on the effect of achievement badges in trakla2 online learning environment. In \textit{Learning and teaching in computing and engineering}, pages 47–54. IEEE.

Hirai et al., 2016. Hirai, M., Muramatsu, Y., Mizuno, S., Kurahashi, N., Kurahashi, H., and Nakamura, M. (2016). Typical visual search performance and atypical gaze behaviors in response to faces in williams syndrome. \textit{Journal of Neurodevelopmental Disorders}, 8.

Khoshnoodifar et al., 2023. Khoshnoodifar, M., Ashouri, A., and Taheri, M. (2023). Effectiveness of gamification in enhancing learning and attitudes: a study of statistics education for health school students. \textit{Journal of Advances in Medical Education \& Professionalism}, 11(4):230.

Laule, 2010. Laule, G. (2010). Positive reinforcement training for laboratory animals. \textit{The UFAW handbook on the care and management of laboratory and other research animals}, pages 206–218.

Licorish et al., 2018. Licorish, S. A., Owen, H. E., Daniel, B., and George, J. L. (2018). Students' perception of kahoot!'s influence on teaching and learning. \textit{Research and Practice in Technology Enhanced Learning}, 13(1):1–23.

Marczewski, 2013. Marczewski, A. (2013). \textit{Gamification: a simple introduction}. Andrzej Marczewski.

Mazurek et al., 2015. Mazurek, A., Bhoopathy, R. M., Read, J. C., Gallagher, P., and Smulders, T. V. (2015). Effects of age on a real-world what-where-when memory task. \textit{Frontiers in Aging Neuroscience}, 7:74.

Murphy et al., 2023. Murphy, D. H., Bjork, R. A., and Bjork, E. L. (2023). Going beyond the spacing effect: Does it matter how time on a task is distributed? \textit{Quarterly Journal of Experimental Psychology}, 76(5):1131–1154.

Qiao et al., 2024. Qiao, S., Yeung, S. S.-s., Shen, X., Leung, J. K. L., Ng, D. T. K., and Chu, S. K. W. (2024). How competitive, cooperative, and collaborative gamification impacts student learning and engagement. \textit{Language Learning \& Technology}.

Rhodes et al., 2019. Rhodes, S., Greene, N. R., and Naveh-Benjamin, M. (2019). Age-related differences in recall and recognition: A meta-analysis. \textit{Psychonomic Bulletin \& Review}, 26(5):1529–1547.

Septiani et al., 2020. Septiani, W. E., Sulistyaningsih, S., and Syakur, A. (2020). The effectiveness of pomodoro technique on students' descriptive text writing quality. \textit{Jurnal Basicedu}, 6(2).

Skinner, 2019. Skinner, B. F. (2019). \textit{The behavior of organisms: An experimental analysis}. BF Skinner Foundation.

Smiderle et al., 2020. Smiderle, R., Rigo, S. J., Marques, L. B., Peçanha de Miranda Coelho, J. A., and Jaques, P. A. (2020). The impact of gamification on students' learning, engagement and behavior based on their personality traits. \textit{Smart Learning Environments}, 7(1):3.

Steel and Klingsieck, 2016. Steel, P. and Klingsieck, K. B. (2016). Academic procrastination: Psychological antecedents revisited. \textit{Australian Psychologist}, 51(1):36–46.

Vlach et al., 2008. Vlach, H. A., Sandhofer, C. M., and Kornell, N. (2008). The spacing effect in children's memory and category induction. \textit{Cognition}, 109(1):163–167.

Wang and Tahir, 2020. Wang, A. I. and Tahir, R. (2020). The effect of using kahoot! for learning–a literature review. \textit{Computers \& Education}, 149:103818.

Yuan, 2022. Yuan, X. (2022). Evidence of the spacing effect and influences on perceptions of learning and science curricula. \textit{Cureus}, 14(1).

Zhang and Yu, 2021. Zhang, Q. and Yu, Z. (2021). A literature review on the influence of kahoot! On learning outcomes, interaction, and collaboration. \textit{Education and Information Technologies}, 26(4):4507–4535.

\newpage
\section{Appendix}
\label{appendix}

The tables below provide details about the participant characteristics and experimental results.

\begin{table}[h!]
\centering
\caption{Descriptive statistics by group for the question "What is your sex?"}
\begin{tabular}{llr}
\toprule
 & What is your sex? & Groups \\
\midrule
N & Male       & 48 \\
  & Female     & 70 \\
\bottomrule
\end{tabular}
\end{table}
\begin{table}[h!]
\centering
\caption{Descriptive statistics by group for the question "What is the highest education level that you have completed?"}
\begin{tabular}{llr}
\toprule
 & Highest completed education level. & Groups \\
\midrule
N & Bachelor's Degree & 12 \\
  & High School       & 98 \\
  & Primary School    &  1 \\
  & Master's Degree   &  5 \\
  & PhD               &  1 \\
  & Secondary School  &  1 \\
\bottomrule
\end{tabular}
\end{table}

\begin{table}[h!]
\centering
\caption{Descriptive statistics by group for the question "What is your major?"}
\begin{tabular}{llr}
\toprule
 & What is your major? & Groups \\
\midrule
N & Faculty of Medicine              & 22 \\
  & Computer Engineering             & 30 \\
  & Psychology                       & 42 \\
  & Law                              &  2 \\
  & Political Sciences               &  1 \\
  & Law and Economy Double Major     &  1 \\
  & Mechatronics Engineering         &  7 \\
  & Industrial Engineering           &  3 \\
  & International Trade and Finance  &  2 \\
  & International Relations          &  2 \\
  & Mechanical Engineering           &  1 \\
  & Electrical-Electronics Engineering &  2 \\
  & Communication Sciences           &  1 \\
  & None                             &  2 \\
\bottomrule
\end{tabular}
\end{table}

\begin{table}[ht]
\centering
\caption{Meaningless Syllables List}
\begin{tabular}{ccccc}
\toprule
Hoz & Tiv & Rev & Yuk & Daf \\
Lem & Gip & Zep & Zud & Tah \\
Lum & Zob & Yob & Taz & Nuv \\
Tey & Fik & Bik & Gos & Zoz \\
\bottomrule
\end{tabular}
\end{table}

\begin{table*}[h]
\centering
\caption{Independent Samples T-Test on \textit{Student's t}}

\begin{tabularx}{\textwidth}{cX|cc|c}
    \hline
    & \textbf{Measure} & \textbf{Reward} & \textbf{No Reward} & \textbf{p} \\
    \hline
    Q1  & It was fun doing the task of the experiment.                 & 3.9 (1.1) & 3.4 (1.0) & 0.019*\\
    Q2  & Rate your performance.                                       & 2.7 (0.8) & 2.7 (0.7) & 0.556 \\
    Q3  & Memorizing syllables was quite tiring.                        & 2.4 (1.2) & 2.6 (1.2) & 0.535 \\
    Q4  & I find it hard to focus on studying.                          & 3.1 (1.2) & 3.3 (0.9) & 0.244 \\
    Q5  & I get bored easily while studying.                            & 3.4 (1.1) & 3.3 (0.9) & 0.729 \\
    Q6  & It is rewarding to watch short videos for fun.                & 3.4 (1.3) & 3.3 (1.3) & 0.666 \\
    Q7  & I give in to temptations while studying easily.               & 3.4 (0.9) & 3.4 (1.0) & 0.681 \\
    Q8  & It was rewarding to watch short videos during the experiment. & 3.2 (1.3) & --        & --    \\
    Q9  & Rate your memory skills.                                      & 3.3 (0.9) & 3.2 (0.8) & 0.417 \\
    Q10 & Found it easy to go back to work after the break.             & 3.3 (1.3) & 3.5 (1.1) & 0.391 \\
    Q11 & The reward system was a positive experience.                  & 3.5 (1.2) & --        & --    \\
    Q12 & Memorizing the syllables was quite boring.                    & 2.5 (1.3) & 2.8 (1.3) & 0.178 \\ \hline
        & Recognition Points                                           & 33.0 (4.4) & 33.3 (4.2) & 0.752 \\
        & Free Recall Points                                           & 8.2 (3.3)  & 7.6 (3.3)  & 0.303 \\
    \hline
\end{tabularx}
\begin{tablenotes}[flushleft] 
    \small
    \item \begin{minipage}{\textwidth}
    The questionnaire assessed multiple dimensions including:
    perceived enjoyment (Q1, Q11), perceived performance (Q2, Q9),
    fatigue and effort (Q3, Q4), boredom (Q5, Q12),
    temptation (Q7), reward valuation (Q6, Q8),
    and persistence (Q10).
    \end{minipage}
    \item \textbf{Note.} Values are reported as mean (standard deviation).
    \item Dashes (--) indicate that one or both groups did not contain enough observations.
    \item Asterisk(*) indicate statistical significant (p$<0.05$)
\end{tablenotes}
\label{tab:t_test}
\end{table*}

\end{document}